\documentclass[aps,prl,twocolumn,showpacs,amsmath,amssymb]{revtex4-1}

\usepackage{graphicx,times}

\begin{document}

\markboth{Igor Omelyan and Yuri Kozitsky}{Spatially inhomogeneous population dynamics: beyond the mean field approximation}

\title{Spatially inhomogeneous population dynamics: beyond the mean field approximation}

\author{Igor Omelyan$^1$ and Yuri Kozitsky$^2$}

\affiliation{$^1$Institute for Condensed Matter Physics, National Academy
of Sciences of Ukraine, 1 Svientsitskii Street, UA-79011 Lviv, Ukraine}
\affiliation{$^2$Instytut Matematyki, Uniwersytet Marii Curie-Sk{\l}odowskiej,
20-031 Lublin, Poland}

\date{\today}

\begin{abstract}

We propose a novel method for numerical modeling of spatially inhomogeneous moment dynamics of populations with nonlocal dispersal and competition in continuous space. It is based on analytically solvable decompositions of the time evolution operator for a coupled set of master equations. This has allowed us -- for the first time in the literature -- to perform moment dynamics simulations of spatially inhomogeneous systems beyond the mean-field approach and to calculate the inhomogeneous pair correlation function using the Kirkwood superposition ansatz. As a result, we revealed a number of new subtle effects, possible in real populations. Namely, for systems with short-range dispersal and mid-range competition, strong clustering of entities at small distances followed by their deep disaggregation at larger separations are observed in the wavefront of density propagation. For populations in which the competition range is much shorter than that of dispersal, the pair correlation function exhibits a long-tail behavior. Remarkably, the latter effect takes place only due to the spatial inhomogeneity and thus was completely unknown before. Moreover, both effects get stronger in the direction of propagation. All these types of behavior are interpreted as a trade-off between the dispersal and competition in the coexistence of reproductive pair correlations and the inhomogeneity of the density of the system.

\vspace{14pt}


\end{abstract}


\maketitle

\section{Introduction}

Population dynamics (PD) is widely studied in mathematical biology, ecology, medicine, and life sciences \cite{Murray, Bolker, Gross, Iannelli}. Many models were proposed during the long history of PD. They include continuum, lattice, network, individual, spatial, and other approaches \cite{Murray, Bolker, Gross}. While the continuum theory is too simplified, the lattice schemes appear to be more accurate \cite{Bolker}. But the lattice representation can modify to some extent real populations where entities take positions continuously in space instead to be located in predefined knots. At the same time, individual-based models (IBMs) yield the most detailed description. However, in numerical simulations, the IBMs may be computationally very expensive, especially for population systems of large sizes \cite{Bolker, South, Lethbridge}.

For overcoming drawbacks of existing PD models, over the last two decades there has been an increasing interest in developing \textit{spatial moment} dynamics (SMD) \cite{BolkerPac, Dieckmann, Law, Murrell, Birch, Adams, Simpson, Plank, Binny, Binnya}. In SMD the populations are described by time-dependent spatial moments (also called correlation functions \cite{FinKonKoz, Ruelle}). The first two of them are the local population density and pair correlation function. The SMD approach can be viewed as an \textit{extension} of the traditional mean-field (MF) theory. The latter is invoked for most PD models to simplify consideration. It should be emphasized that MF totally neglects the second-, third- and higher-order spatial correlations. On the other hand, in SMD these correlations are explicitly accounted.

The SMD models were applied to ecological dynamics, spatial epidemics, surface chemistry reactions, predator-prey metapopulations (see \cite{Bolker, Simpson} and the references therein). These models are particularly useful in detecting patchiness and clustering \cite{Law, Young} in the spatial distribution of different organisms, such as trees in a beech forest \cite{LawIll} or breast cancer cells at an in vitro growth-to-confluence assay \cite{Agnew}. Strictly speaking, the SMD approach is able to predict subtle effects which are unreachable within the MF framework. The former can also be employed to improve or revisit some MF data, e.g., on the formation of patterns in evolution of bacterial colonies \cite{Fuentes}.

The SMD description is exact if the infinite number of spatial moments is involved in the hierarchy of the master equations. For practical reasons, this hierarchy needs to be closed, since in computer simulations we cannot operate with the infinite number of equations. Usually the closure is performed at the third-order level, so that numerical solutions are found for the first two equations (which cannot be handled analytically). This is indubitably \textit{superior} to the MF approximation and can provide \cite{Dieckmann, Binny, Binnya} a high accuracy (comparable to that of the IBMs) for observables such as population density and pair correlation function. Several closures of powers from one to three have been introduced for SMD \cite{Dieckmann, Murrell, Binny, Binnya}. It was realized that their precision increases with increasing the power number.

Despite the mentioned achievements, all the previous SMD simulations of continuous-space models were restricted exclusively to the spatially \textit{homogeneous} case \cite{BolkerPac, Dieckmann, Law, Murrell, Adams, Binny, Binnya}. Obviously, this presents a significant limitation as then most of the principal properties of population systems are inaccessible. For example, the account of spatial inhomogeneity is essential in the study of the wavefront and spread dynamics. Inhomogeneous processes are important in ecological invasions, in vitro cell invasion assays, embryogenesis and wound healing, malignant tumor proliferation, etc (see \cite{Plank} and the references therein). All these processes involve colonisation of a region by a population of agents that is initially spatially confined. As was underlined in Ref.~\cite{Plank}, very little is known about SMD for inhomogeneous systems.

Until now, there have been \textit{no} publications on continuous-space \textit{inhomogeneous} SMD (ISMD) simulations. This is explained by the fact (carefully inspected in this work) that the standard numerical methods are incapable for ISMD. Thus, the main goals of this work are: (i) development of an approach \textit{enabling} to solve the problem with spatial inhomogeneity; (ii) carrying out \textit{first} ISMD simulations; and (iii) discovering \textit{new} PD effects.

\section{Model}

Consider a population of point entities dwelling in continuous space $\mathbb{R}^d$. The entities reproduce themselves, disperse, and die on their own or due to competition. Let $n_t(x)$, $u_t(x,y)$, and $w_t(x,y,z)$ be the spatial moments of the first, second, and third order, respectively. They determine the probability density of finding at time $t$ a single entity in coordinate point $x \in \mathbb{R}^d$, a pair of entities in $x,y$, or a triplet in $x,y,z$. Then the first two coupled integro-differential equations of the ISMD hierarchy \cite{Birch, Plank, Binny, Binnya, FinKonKoz} can be written as
\begin{eqnarray} \label{1}
\frac{d n_t(x)}{d t} = &&\ \!\int\! d y \big[ a(x,y) n_t(y) - b(x,y) u_t(x,y) \big] - m n_t(x) , \nonumber \\[-2pt] \\[-2pt] \label{2} \frac{d u_t(x,y)}{d t} &&\ = a(x,y) \big( n_t(x) + n_t(y) \big) - 2 b(x,y) u_t(x,y) \nonumber \\ &&\ + \!\int\! d z \big[ a(x,z) u_t(y,z) + a(y,z) u_t(x,z) \\ [2pt] &&\ - \ \big( b(x,z) + b(y,z) \big) w_t(x,y,z) \big] - 2 m u_t(x,y) . \nonumber
\end{eqnarray}

Here $a(x,y)$ and $b(x,y)$ define the probabilities per unit time for dispersal to point $x$ of an entity born ($+$) at $y$ and its death ($-$) in $x$ caused by competition with a neighbor at $y$, while $m$ is the intrinsic mortality. The kernels $a$ and $b$ are modeled by the Gaussians $c_\pm/(2 \pi \sigma_\pm^2)^{d/2} \exp[-(x-y)^2/(2 \sigma_\pm^2)]$ or the top-hat functions $c_\pm/(2 \sigma_\pm)^d$ for $|x-y| \le \sigma_\pm$ with the intensities $c_\pm$ and ranges $\sigma_\pm$ of dimensionality $d=$ 1, 2, or 3. Note that the dispersal and competition kernels are normalized so that $\int\! a(x,y) d y \!=\! c_+$ and $\int b(x,y) d y \!=\! c_-$. The most general form of the ISMD equations (which include motility and mutation of entities) is presented in Ref.~\cite{Plank}.

Although more complex ISMD models can be introduced, too \cite{Plank, Binny, Binnya}, Eqs.~(\ref{1}) and (\ref{2}) are quite complicated. In simplified limits, equation (\ref{1}) transforms to the well-known equations of previous spatial PD models. For example, neglecting pair correlations within the MF approximation by putting $u_t(x,y) \simeq n_t(x) n_t(y)$, we come from Eq.~(\ref{1}) to the kinetic equation of Ref.~\cite{FinKonKoz}. Additionally, by letting $\sigma_+ \ll 1$ (local dispersal) we have $\!\int\! a(x,y) n_t(y) d y \approx c_+ n_t(x) + D \partial^2 n_t(x)/\partial x^2$ that leads to the diffusion MF model \cite{Fuentes}, where $D = c_+ \sigma_+^2/2$ is the diffusion coefficient. Finally, in the limit $\sigma_- \to +0$ of local competition when $b(x,y) \to c_- \delta(x-y)$, we reproduce from Eq.~(\ref{1}) the classical Fisher-Kolmogorov-Petrovsky-Piscunov \cite{Fisher, Kolomogor, Maruvka} reaction-diffusion equation $d n_t(x)/d t = (c_+ - m) n_t(x) - c_- n_t^2(x) + D \partial^2 n_t(x)/\partial x^2$, intensively exploited in early investigations.

Despite the importance of the ISMD equations (\ref{1}) and (\ref{2}), there were no successful attempts reported to find their solutions numerically (they cannot be obtained analytically). As will be shown later, the reason is that the existing numerical methods are inappropriate to solve these equations in the case of inhomogeneous conditions. That is why the assumption of spatial homogeneity, i.e. that $n_t(x)$ does not change on coordinate $x$, was made \cite{BolkerPac, Dieckmann, Murrell, Binny, Binnya}. Then, the second-order function $u_t(x,y)$ will depend only on the difference $|x-y|$ and not on $x$ and $y$ separately, significantly simplifying the computations and enabling to obtaining spatially homogeneous results. It was mentioned in the Introduction that the homogeneous approach is very restrictive in comparison with ISMD. By ignoring density variations, similar simplifications were used when incorporating spatial correlations into lattice models \cite{Baker, Markham, Markhama}.

First studies on improving the MF approach by including inhomogeneous correlations were carried out in Ref.~\cite{Simpsona}. However, the consideration was devoted solely to lattice models within a nearest-neighbor scheme in terms of the average site occupancy probability. Efforts to estimate the inhomogeneous pair correlation function as a weighted sum of its homogeneous counterparts corresponding to different constant densities treated as local ones were also made \cite{LawIll}. Our method derived below is grounded on the theoretically rigorous framework developed for continuous-space ISMD models in the presence of spatial inhomogeneity, where homogeneous conditions appear as a particular case.

\section{Method}

The main concepts of our ISMD approach consist in the following. Firstly, in order to solve Eqs.~(\ref{1}) and (\ref{2}) we perform their discretization using the equalities
\begin{eqnarray} \label{3}
\frac{d n_i}{d t} &=& h \sum_j \big( a_{ij} n_j - b_{ij} u_{ij} \big) - m n_i , \\ \label{4} \frac{d u_{ij}}{d t} &=& a_{ij} \big( n_i+n_j \big) - 2 \big( m + b_{ij} \big) u_{ij} \nonumber \\ &+& h \sum_k \big[ a_{ik} u_{jk} + a_{jk} u_{ik} - ( b_{ik} + b_{jk}) w_{ijk} \big] .
\end{eqnarray}
Here the sums represent the spatial integrals over $y$ and $z$, while $n_i(t) = n_t(x_i)$, $u_{ij}(t) = u_t(x_i,x_j)$, and $w_{ijk}(t) = w_t(x_i,x_j,x_k)$ are the values of the correlation functions in grid points $x_{i,j,k}$ uniformly distributed inside the region $[-L/2,L/2]^d$ with spacing $h=(L/N)^d$, and $i,j,k = 1, 2, \ldots, N^d$. The kernel values in the grid points are denoted by $a_{ij} = a(x_i, x_j)$ and $b_{ij} = b(x_i, x_j)$. Area $[-L/2,L/2]^d$ constitutes an interval, a square, or a cube in the cases $d=1$, $2$, or $3$, respectively.

Note that the length $L$ should be sufficiently long with respect to all characteristic coordinate scales of the population system. Number $N$ of grid points must be large enough to minimize the noise caused by the discretization. Then mesh $h$ will be sufficiently small to provide a high accuracy of the spatial integration. Obviously, in the limits $L, N \to \infty$ and $h \to 0$, the discretized Eqs.~(\ref{3}) and (\ref{4}) coincide with their original, continuous counterparts [Eqs.~(\ref{1}) and (\ref{2})]. The finite-size effects can be reduced by employing the corresponding boundary conditions when mapping our infinite range $x,y,z \in ]-\infty, \infty[^d$ by the finite area $x_i, x_j, x_k \in [-L/2,L/2]^d$. If the entities initially ($t=0$) exist only within a narrow region $[-l/2,l/2]$ with $l \ll L$, and they are absent outside of it, i.e., $\{ n_0(x), u_0(x,y) \}\big|_{|x|,|y| > l/2} = 0$, we can apply the Dirichlet boundary conditions $\lim_{x,y \to \pm \infty} \{ n_t(x), u_t(x,y) \} = 0$. This means that nonzero values of the spatial moments will not reach the area boundaries $\pm L/2$ in each direction during the simulations over the finite simulation time $0 < t \le T$. When $n_0(x)$ and $u_0(x,y)$ take nonzero values anywhere in infinite space, it is necessary to use the periodic boundary conditions.

Secondly, to decouple the ISMD hierarchy, we apply the power-3 closure \cite{Dieckmann, Murrell, Binny, Binnya}
\begin{equation} \label{4a}
w_t(x,y,z) = \frac{u_t(x,y) u_t(x,z) u_t(y,z)}{n_t(x) n_t(y) n_t(z)}
\end{equation}
for the third-order correlation function. This closure is well known \cite{Kirkwood, Hansen, BenNaim} in theoretical physics as the Kirkwood superposition approximation (KSA). For the discretized dublicate of $w_t(x,y,z)$ Eq.~(\ref{4a}) yields $w_{ijk} = u_{ij} u_{ik} u_{jk} / ({n_i n_j n_k})$. We see that in the KSA ansatz, the moment $w$ of the highest order is expressed in terms of the lower-order correlation functions $u$ and $n$. As was shown earlier in the spatially homogeneous case, this ansatz provides much better accuracy when reproducing the third-order correlations than the so-called power-2 and -1 closures \cite{Dieckmann, Murrell, Binny, Binnya}.

Thirdly, let us introduce the set $\Gamma=\{n_i, u_{ij}\}$ of dynamical variables. Then the complicated coupled system of $N^d+N^d \times N^d$ autonomous ordinary differential equations (\ref{3}) and (\ref{4}) with respect to the same number of unknown quantities $n_i$ and $u_{ij}$, where $i,j=1,2,\ldots,N^d$, can be cast in the compact Liouville form
\begin{equation} \label{4b}
\\ [-12pt]
\frac{d \Gamma}{d t} = \Psi \Gamma ,
\\ [2pt]
\end{equation}
where
\begin{equation} \label{4c}
\Psi = \sum_{i=1}^{N^d} \Psi_i + \sum_{i \le j}^{N^d} \Psi_{ij}
\end{equation}
is the differential operator. Its components are $\Psi_{ij} = \Psi_{ij}^{\rm I} + \Psi_{ij}^{\rm II}$ and
\vspace{-4pt}
\begin{align} \label{5}
\begin{split}
\Psi_i & = \Big( \alpha_i - \beta_i n_i \Big) \frac{\partial}{\partial n_i} , \\ \Psi_{ij}^{\rm I} & = \Big( \alpha_{ij} - \beta_{ij} u_{ij} \Big)
\frac{\partial}{\partial u_{ij}} , \\
\Psi_{ij}^{\rm II} & = - \Big( \big(1-\delta_{ij}\big) \gamma_{ij} u_{ij}^2 +
\delta_{ij} \zeta_{ii} u_{ii}^3 \Big) \frac{\partial}{\partial u_{ij}} .
\end{split}
\end{align}
The decomposition coefficients entering to Eq.~(\ref{5}) can be expressed as
\begin{eqnarray} \label{6}
\alpha_i &=& h \sum_{j \ne i} a_{ij} n_j - h \sum_j b_{ij} u_{ij} ,
\ \ \ \ \beta_i = m - h a_{ii} , \\ [7pt] \label{7}
\alpha_{ij} &=& a_{ij} \big( n_i + n_j \big) + h \sum_{k \ne i} a_{ik} u_{jk}
+ h \sum_{k \ne j} a_{jk} u_{ik} , \nonumber \\ [1pt]
\beta_{ij} &=& 2 \big( \beta_i + b_{ij} \big) + h \sum_{k \ne i,j} \Big( b_{ik} +
b_{jk} \Big) \frac{u_{ik} u_{jk}}{n_i n_j n_k} , \\ [1pt]
\gamma_{ij} &=& h \frac{b_{ii} + b_{ij}}{n_i n_j}
\Big( \frac{u_{ii}}{n_i} + \frac{u_{jj}}{n_j} \Big) , \ \ \ \
\zeta_{ii} = \frac{2 h b_{ii}}{n_i^3} . \nonumber
\end{eqnarray}

Fourthly, in view of Eq.~(\ref{4b}), the solution to the ISMD equations is
\begin{equation} \label{7a}
\Gamma(t) = \left[e^{\Psi \Delta t}\right]^K \Gamma(0) ,
\end{equation}
where $\Delta t$ and $K=t/\Delta t$ denote the time increment and total number of steps, respectively. Since the time evolution operator $e^{\Psi \Delta t}$ cannot be handled exactly, proceeding in the spirit of Refs.~\cite{Omelyans} and \cite{Omelyana} we derive the multistage \textit{decomposition} propagation (DP):
\begin{equation} \label{8}
e^{\Psi \Delta t} = \prod_{i,j=1}^{N^d}\! e^{\Psi_{ij} \frac{\Delta t}{2}} \prod_{i=N^d}^1 \ \prod_{i=1}^{N^d} e^{\Psi_i \frac{\Delta t}{2}}
\!\prod_{i,j=N^d}^1\! e^{\Psi_{ij} \frac{\Delta t}{2}} + {\cal O}(\Delta t^3) .
\end{equation}
Here the factorization is performed symmetrically with respect to $i$ and $i,j$, while $i \le j$ because of $u_t(y,x) = u_t(x,y)$, i.e., $u_{ji}=u_{ij}$. Due to the specially tailored decomposition [Eq.~(\ref{5})] of $\Psi$, each of the single exponentials appearing in Eq.~(\ref{8}) can be evaluated \textit{analytically}. Indeed, the coefficients $\alpha_i$ and $\beta_i$ do not depend on $n_i$ for every $i$, as this follows from Eq.~(\ref{6}). Then according to Eq.~(\ref{5}), the operator $e^{\Psi_i \Delta t/2}$ acting on the local density $n_i$ results in the analytical solution
\begin{equation} \label{9}
e^{\Psi_i \frac{\Delta t}{2}} n_i = n_i e^{-\beta_i \Delta t/2} +
\big( 1 - e^{-\beta_i \Delta t/2} \big) \alpha_i \big/\beta_i .
\end{equation}
Moreover,
\begin{equation} \label{9a}
e^{\Psi_{ij} \frac{\Delta t}{2}} = e^{\Psi_{ij}^{\rm II} \frac{\Delta t}{4}} e^{\Psi_{ij}^{\rm I} \frac{\Delta t}{2}} e^{\Psi_{ij}^{\rm II} \frac{\Delta t}{4}} + {\cal O}(\Delta t^3)
\end{equation}
with
\begin{eqnarray} \label{10}
e^{\Psi_{ij}^{\rm I} \frac{\Delta t}{2}} u_{ij} &=& u_{ij} e^{-\beta_{ij} \Delta t/2}
+ \big( 1 - e^{-\beta_{ij} \Delta t/2} \big) \alpha_{ij} \big/ \beta_{ij} , \nonumber \\ [4pt] e^{\Psi_{ij}^{\rm II} \frac{\Delta t}{4}} u_{ij} &=& u_{ij}\big/
\big( 1+\gamma_{ij} u_{ij} \Delta t/4 \big) , \ \ \ \text{for}\ \ i \ne j , \\ [3pt]
e^{\Psi_{ii}^{\rm II} \frac{\Delta t}{4}} u_{ii} &=& u_{ii}\big/
\big( 1+\zeta_{ii} u_{ii}^2 \Delta t/2 \big)^{1/2} , \nonumber
\end{eqnarray}
where Eq.~(\ref{5}) and the independence of $\{\alpha,\beta,\gamma\}_{ij}$ or $\zeta_{ii}$ on $u_{ij}$ or $u_{ii}$ at given $i$ and $j$ [see Eq.~(\ref{7})] have been used.

In such a way, the numerical solutions $n_i(t)$ and $u_{ij}(t)$ are obtained for any time $0 < t \le T$ by consecutively applying Eqs.~(\ref{7a})--(\ref{10}). Of course, DP given by Eq.~(\ref{8}) is not exact, so that ${\cal O}(\Delta t^3)$-uncertainties arise. However, they can be reduced to an arbitrary small level by decreasing the size $\Delta t$ of the time step.

\section{Results}

The ISMD/KSA/DP simulations were carried out in $d=1$ at $L=80$ and $h=0.0125$ with $N=6400$ and $\Delta t=0.05$. Further increasing space and time resolution does not affect the solutions. The initial ($t=0$) density distribution $n_0(x)$ was the Gaussian centered at $x=0$ with $c_0=1$ and $\sigma_0=1$. Then $n_t(-x)=n_t(x)$, and thus $n_t(x)$ will be presented only for $x \ge 0$. The Dirichlet boundary conditions were used to exclude the finite-size effects. Since we have five parameters ($m$, $c\pm$, and $\sigma_\pm$) of the model, Eqs.~(\ref{1}) and (\ref{2}) can describe various systems in different areas. We consider two characteristic examples. The first one is a system (of type 1) with short-range dispersal, $\sigma_+=0.1$, and mid-range competition, $\sigma_-=1$, modeled by the top-hat kernels. The second example (type 2) concerns short-range competition, $\sigma_-=0.1$, and mid-range dispersal, $\sigma_+=1$, for the Gaussians. For both types, a small mortality, $m=0.01$, and moderate intensities, $c_\pm=1$, were supposed.

\begin{figure}
\includegraphics[width=86mm]{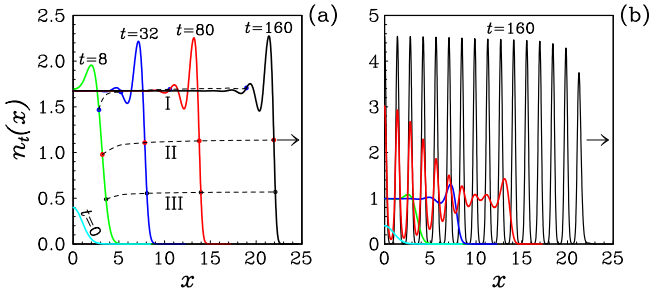}
\caption{Propagation of density distribution obtained by the ISMD [subset (a)] and MF [subset (b)] models with short-range dispersal.}
\label{f1}
\end{figure}

\begin{figure}
\includegraphics[width=86mm]{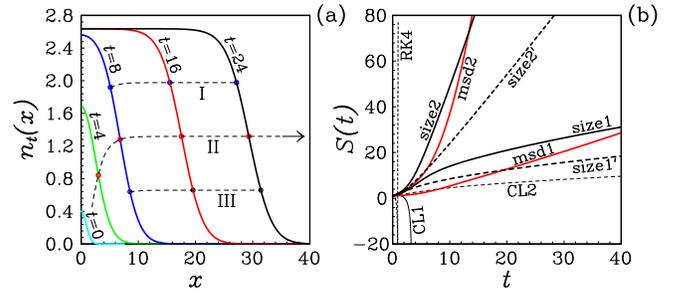}
\caption{(a) The same as for Fig.~\ref{f1}a but for short-range competition; (b) The ISMD (solid curves) and MF (dashed curves) population size $S(t)$ and mean square displacement of entities for systems 1 and 2.}
\label{f2}
\end{figure}

The ISMD/KSA/DP densities $n_t(x)$ are shown in Figs.~\ref{f1}a (type 1) and \ref{f2}a (type 2). The MF data (type 1) are presented in Fig.~\ref{f1}b. From Fig.~\ref{f1} one can see that the MF approximation incorrectly predicts a periodic structure with deep amplitude modulation in a steady state at $t \gtrsim 160$. On the other hand, \textit{no} such pattern arises within the accurate ISMD description. Here, with increasing $t$, the function $n_t(x)$ becomes flat in $x$ near $x=0$, while a strong oscillating-like inhomogeneity is maintained at the wavefront of the propagation. This striking difference is a consequence of the MF assumption $u_t(x,y) \equiv n_t(x) n_t(y)$ in which any spatial correlations are completely neglected altogether. The fact that the MF approach can fail \textit{dramatically} in some cases was mentioned earlier for lattice models \cite{Baker, Markham, Markhama}. For type 2, the ISMD density profiles are more smooth (cf. Figs.~\ref{f2}a and \ref{f1}a) and similar in shape with the MF ones (not shown) but noticeably larger than the latter in amplitude.

The total number $S(t) = \!\int\! n_t(x) d x$ of entities and their mean square displacement $\langle x^2 \rangle(t) = \!\int\! x^2 n_t(x) d x/S(t)$ are plotted in Fig.~\ref{f2}b versus $t$. We see that the MF model appreciably underestimates values of $S(t)$ for both the types. The ISMD function $\langle x^2 \rangle$ of type 1 begins to depend linearly on $t$ in a steady state after a relaxation time of $t \sim 32$. The linearity $\langle x^2 \rangle \sim t$ indicates about a diffusive-like behavior inherent to local dispersal ($\sigma_+ \ll 1$). For type 2, we have $\langle x^2 \rangle \sim t^2$ in a steady state at $t \gtrsim 24$, meaning that a regular regime with $S \sim t$ takes place.

\begin{figure}[b]
\includegraphics[width=86mm]{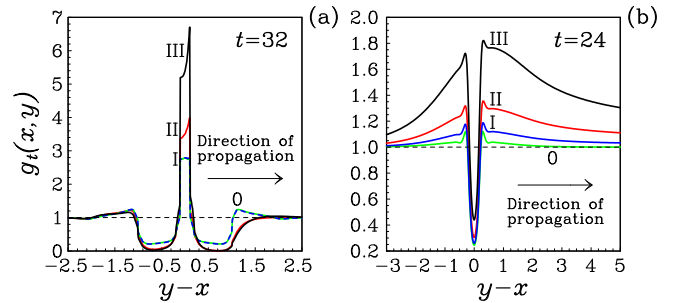}
\caption{The ISMD inhomogeneous pair correlation functions at $x=0$ and in the wavefront region ($x=x_{\rm I,II,III}$) for short-range dispersal [subset (a)] and short-range competition [subset (b)].}
\label{f3}
\end{figure}

The inhomogeneous pair correlation functions $g_t(x,y) = u_t(x,y) \big/ \big(n_t(x) n_t(y)\big)$ are presented in Fig.~\ref{f3} at $t=32$ (type 1) and $t=24$ (type 2) as dependent on $y-x$ for $x=0$ and three wavefront points $x=x_{\rm I,II,III}$. The latter were chosen such that $n_t(x_{\rm I,II,III})$ decreases to the levels $3/4$, $1/2$, and $1/4$ with respect the maximum of $n_t$ (see circles connected by dashed curves in Figs.~\ref{f1}a and \ref{f2}a). Fig.~\ref{f3} demonstrates that $g_t(x,y)$ can deviate \textit{significantly} from the MF value 1. Note that the initial condition $g_0(x,y) = 1$ with no pair correlations was utilized at $t=0$. These correlations are quickly reproduced owing to the interactions, so that already at $t \gtrsim 32$ (type 1) or $t \gtrsim 24$ (type 2) we achieve the steady states.

For type 1, we observe a strong \textit{clustering}, $g_t(x,y) \gg 1$, of entities in the narrow interval $|y-x| \lesssim \sigma_+$ at the wavefront ($x=x_{\rm I,II,III}$) of density propagation, see Fig.~\ref{f3}a. With increasing distance $|y-x|$, the clustering suddenly transforms into a wide area of deep \textit{disaggregation}, $g_t \approx 0$. In the homogeneous domain ($x = 0$) these effects are not so visible. Note also that $g_t(0,-y)=g_t(0,y)$, whereas $g_t(x,y)$ is an asymmetric function in $y-x$ at $x \ne 0$. For type 2, the IPCF identifies an intense \textit{disaggregation} at small separations $|y-x| \sim 0$, where $g_t(x,y) \ll 1$ (look at Fig.~\ref{f3}b). Near $|y-x| \gtrsim \sigma_-$, the disaggregation changes to a moderate \textit{clustering} ($1 < g_t < 2$). At the wavefront ($x=x_{\rm I,II,III}$), the spatial correlations are maintained up to long distances $|y-x| \sim 10-20$, where $g_t(x,y)$ decreases to its asymptotic value 1 very slowly with increasing $|y-x|$. This effect becomes \textit{stronger} in the direction of propagation. \textit{No} such \textit{long tails} are detected in the homogeneous region $x=0$, where $g_t(x,y) \to 1$ already at $|y-x| \gtrsim 3$, just as in Fig.~\ref{f3}a for type 1 at any $x$.

\section{Discussion and Conclusions}

The above effects can be explained by a subtle \textit{interplay} between the dispersal and competition forces in the presence of reproductive pair correlations at inhomogeneous density distributions. Indeed, as the distance over which offspring disperse is made shorter (by reducing $\sigma_+$), individuals are increasingly \textit{clustered} in space, $g_t \gg 1$, around points where they were born. That small portion of entities which has dispersed outside the narrow interval $|x-y|<\sigma_+$ is soon killed, $g_t \approx 0$, by the neighboring agents owing to the competition with them in the wide domain $\sigma_-<|x-y|<\sigma_+$. This leads to deep \textit{disaggregation}, the pattern observed in Fig.~\ref{f3}a, where $\sigma_+ \ll \sigma_-$. In the opposite regime $\sigma_- \ll \sigma_+$, the competition interactions acting over the narrow interval $|x-y|<\sigma_-$ are local and strong. As a result, a sizeable part of agents in this interval dies immediately after their birth, $0< g_t \ll 1$, while survivors are overdispersed up to \textit{long} distances $|x-y| \gtrsim \sigma_+$ with moderate correlations $1 < g_t < 2$. This picture is seen in Fig.~\ref{f3}b, where $\sigma_+ \gg \sigma_-$.

The two types of behavior just described are visible to some extent even in the homogeneous zone (0-curves of Fig.~\ref{f3}). In the inhomogeneous region, they become much more \textit{evident} (I,II,III-curves), especially in the direction of density propagation. This somewhat unexpected behavior can be treated as follows. At the wavefront, the local density $n_t(x)$ rapidly decreases to zero with increasing $|x|$. Then the relative contribution of the term $a(x,y) \big( n_t(x) + n_t(y) \big)$ in the rhs of the first line of Eq.~(\ref{2}) grows. It describes dispersal of the daughter cell to $x$ from the parent at $y$ and vice versa, and thus is responsible for \textit{reproductive} pair correlations (RPCs). This term is proportional to $n_t$, while all others are weighted by $n_t^2$ or $n_t^3$ since $u_t(x,y) = g_t(x,y) n_t(x) n_t(y)$ and $w_t \sim n_t^3$. At small $n_t$ this means that the RPC processes are dominant over the competition ones, leading to an increase of $g_t$. For the same reason, the asymmetric \textit{long tails} appear at $\sigma_+ \gg \sigma_-$ in the inhomogeneous regions, as the correlations are strong enough ($1 < g_t < 2$) at $|x-y| \sim \sigma_+$ and cannot quickly disappear with increasing $|x-y|$ due to the RPCs. \textit{No} long tails arise at $\sigma_+ \ll \sigma_-$ because of the wide-range deep disaggregation in this case. They are also \textit{absent} in the homogeneous regions where the relative impact of the RPCs is small.

It should be emphasized that the RPCs are a \textit{uniquely} biological complication with \textit{no} analogue in the physical and chemical problems \cite{Young}. The most conspicuous example is from physics of liquids where $g_t(x,y)$ tends to 1 in the limit $n_t \to +0$ of small densities \cite{Hansen}. In our case, this function can take arbitrary positive values at $n_t \to +0$ unless $|x-y| \to \infty$. The reproduction of entities is a compelling reason \cite{Birch} for the failure of the MF (Poisson) assumption, $g_t(x,y) = 1$, even at $n_t \to +0$. Remember that $g_t(x,y)$ is the probability of finding one entity at $x$ and another one at $y$ relative to the probability of having entities at $x$ and $y$ if they were independently distributed. Any real organisms are born next to their siblings. Therefore, reproduction ineluctably creates non-Poisson spatial correlations, $g_t \ne 1$, between individuals (daughter cells).

\begin{figure}
\includegraphics[width=86mm]{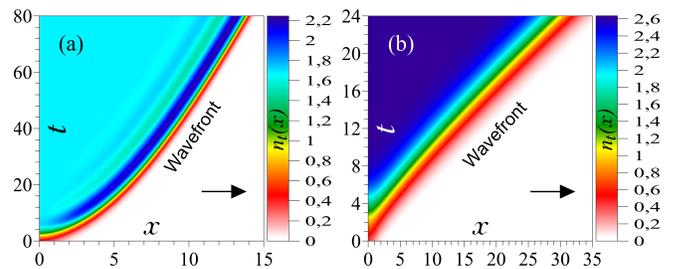}
\caption{The ISMD wavefront dynamics for short-range dispersal [subset (a)] and short-range competition [subset (b)].}
\label{f4}
\end{figure}

The wavefront dynamics is displayed in Fig.~\ref{f4} using the continuous time-space representation for $n_t(x)$. Note that having $n_t(x)$, the spreading speed can be calculated as $v_t(x)=-(d n_t(x)/d t)/(\partial n_t(x)/\partial x)$. We see that the shape and curvature of the wavefront are quite different in the two cases. As for the inhomogeneous pair correlation functions, this is caused by the different types of the interference between the dispersal, competition, and RPCS at spatial inhomogeneity. Similar results to those presented in Figs.~\ref{f1}--\ref{f4} for $d=1$ were observed at $d=2$. They will be presented and discussed in a separate paper elsewhere.

Our investigations have shown that the previously known standard methods working well at spatially homogeneous conditions are \textit{unsuitable} for solving the ISMD equations. As an illustration, in Fig.~\ref{f2}b we include the data obtained for $S(t)$ by the classical Runge-Kutta (RK4) scheme of the fourth order (it is commonly exploited \cite{Baker, Markham, Markhama} in spatially homogeneous PD). It can be seen (RK4-curve) that $S(t)$ quickly becomes negative and unstable with huge deviations even though a very small time step of $\Delta t = 10^{-4}$ is employed. As was realized, the same drawback is inherent to all other standard numerical methods. The difficulty with these methods in the presence of spatial inhomogeneity can be explained by the existence of a \textit{singularity} in the KSA third-order correlation function $w_t(x,y,z) = u_t(x,y) u_t(x,z) u_t(y,z)\big/\big(n_t(x) n_t(y) n_t(z)\big)$ for regions where $n_t$ is close to zero. The standard methods cannot handle this singularity because they are built on regular finite-difference schemes. In the ISMD/KSA/DP approach, the above singularity is \textit{removed} since this part of the dynamics is integrated \textit{analytically} by the product (\ref{8}) of exponential transformations (\ref{9}) and (\ref{10}), guarantying the positiveness of the spatial moments. For instance, the rhs of the equalities in Eq.~(\ref{10}) always remains positive by construction for any $\beta$ owing to the fact that $\alpha$, $\gamma$, and $\zeta$ are greater than zero according to Eq.~(\ref{7}). As these equalities are exact, they can be applied at any values of $\alpha$, $\beta$, $\gamma$, $\zeta$, and $\Delta t > 0$. The RK4 method fails in the singular region, where $\gamma$ and $\zeta$ can be large due to the RPCs, because then the conditions $\gamma u \Delta t/4 \ll 1$ and $\zeta u^2 \Delta t/2 \ll 1$ are violated at normal sizes of $\Delta t$. These conditions are mandatory for RK4 but not for DP in view of the analyticity of the latter.

The positiveness is also provided by the third-order KSA. Closures of lower orders cannot ensure positive solutions and thus are not appropriate. The functions $S(t)$ obtained within the power-1 and -2 closures \cite{Dieckmann, Murrell} are included in Fig.~\ref{f2}b as well (curves marked by CL1 and CL2) for type 1. We see that the CL2 scheme produces the results even worse than those of the MF approximation, while the CL1 curve falls into the negative region (the same behavior was observed for type 2).

The limitations of the ISMD/KSA/DP approach are caused by the approximate character of the KSA closure \cite{Hansen, BenNaim}. As a consequence, it cannot be used at those values of the ISMD model parameters ($c_\pm$, $\sigma_\pm$, $m$) which lead to critical regimes or to $(x,y$)-regions where $g_t(x,y)$ is extremely high. The exact closure can be represented as an infinite diagrammatic series in terms of multiple integrals of correlation functions \cite{Hansen}. However, these integrals are cumbersome and computationally intractable. Similar challenges, including effects of short loops in graph topologies, arise within adaptive network models \cite{Gross, Smaldino, Demirel}. A way to improve the ISMD/KSA/DP method consists in adding the third equation \cite{Plank} for the fourth-order spatial moment to Eqs.~(\ref{1}) and (\ref{2}), complemented by the (more precise) Fisher-Kopelovich closure \cite{Somani}. All these questions will be the subject of our future researches.

The DP technique developed in this paper can be considered as the first extension of the powerful decomposition methodology \cite{Omelyans, Omelyana} widely exploited in molecular dynamics simulations of liquids to the field of population dynamics. Its powerfulness is provided by the preservation of characteristics features inherent in exact solutions, such as reversibility and symplecticity of flow in phase space for liquids or positiveness of distributions for population systems. Other techniques such as the Liouville formalism, concepts of dynamical variables, hierarchies for master equations, and closure schemes, all taken from non-equilibrium statistical physics of liquids, were also used in our work.

\section{Summary}

We have derived a \textit{novel approach} to population dynamics simulations of spatially \textit{inhomogeneous} birth-death systems with nonlocal dispersal and competition. It is based on the DP technique to solve numerically the master equations for spatial moments of entity distribution in continuous space by splitting the time evolution operator into analytically solvable parts. This has enabled us to perform the \textit{first} ISMD modeling, as well as to find and explain \textit{new} subtle effects which can take place in real populations. They include the possible presence of asymmetric long tails in inhomogeneous pair correlation functions, as well as the coexistence of strong clustering and deep disaggregation in the wavefront of spatial propagation.

The proposed approach can readily be adapted to more complex, multicomponent ISMD models \cite{Plank, Binny, Binnya} by including neighbor-dependent birth, motility, marked agents, mutation, directionally biased movement, etc. The ISMD simulations can be expanded to higher dimensions. The corresponding results on these topics will be presented in a separate publication.

\section*{Acknowledgment}

Part of this work was supported in 2017 by the European Commission under the project STREVCOMS PIRSES-2013-612669. Y. K. acknowledges the support in 2018 of the National Science Centre, Poland, grant 2017/25/B/ST1/00051.


\begin{thebibliography}{99}


\bibitem{Murray}
J. D. Murray, \textit{Mathematical Biology} (Springer-Verlag, New York, 2002). 

\bibitem{Bolker}
B. M. Bolker, Continuous-space models for population dynamics, in \textit{Ecology, Genetics, and Evolution in Metapopulations}, eds. I. Hanski and O. Gaggiotti (Academic, NewYork, 2004), pp. 45--69. 

\bibitem{Gross}
T. Gross and H. Sayama, \textit{Adaptive Networks: Theory, Models and Applications} (Springer-Verlag, New York, 2009). 

\bibitem{Iannelli}
M. Iannelli and A. Pugliese, \textit{An Introduction to Mathematical Population Dynamics} (Springer, New Delhi, 2014). 


\bibitem{South}
A. South, Dispersal in spatially explicit population models, \textit{Conservation Biology} \textbf{13} (1999) 1039--1046. 

\bibitem{Lethbridge}
M. R. Lethbridge and J. C. Strauss, A novel dispersal algorithm in individual-based, spatially-explicit Population Viability Analysis: A new role for genetic measures in model testing?, \textit{Environ. Model. Softw.} \textbf{68} (2015) 83--97. 


\bibitem{BolkerPac}
B. Bolker and S. W. Pacala, Using moment equations to understand stochastically driven spatial pattern formation in ecological systems, \textit{Theor. Popul. Biol.} \textbf{52} (1997) 179--197. 

\bibitem{Dieckmann}
U. Dieckmann and R. Law, Relaxation projections and the method of moments, in \textit{The Geometry of Ecological Interactions: Simplifying Spatial Complexity}, eds. U. Dieckmann, R. Law and J.A.J. Metz (Cambridge University Press, Cambridge, 2000), pp. 412--455. 

\bibitem{Law}
R. Law, D. J. Murrell and U. Dieckmann, Population growth in space and time: spatial logistic equations, \textit{Ecology} \textbf{84} (2003) 252--262. 

\bibitem{Murrell}
D. J. Murrell, U. Dieckmann and R. Law, On moment closures for population dynamics in continuous space, \textit{J. Theor. Biol.} \textbf{229} (2004) 421--432. 

\bibitem{Birch}
D. A. Birch and W. R. Young, A master equation for a spatial population model with pair interactions, \textit{Theor. Popul. Biol.} \textbf{70} (2006) 26--42. 

\bibitem{Adams}
T. P. Adams, E. P. Holland, R. Law, M. J. Plank and M. Raghib, On the growth of locally interacting plants: differential equations for the dynamics of spatial moments, \textit{Ecology} \textbf{94} (2013) 2732--2743. 

\bibitem{Simpson}
M. J. Simpson and R. E. Baker, Special issue on spatial moment techniques for modelling biological processes, \textit{Bull. Math. Biol.} \textbf{77} (2015) 581--585. 

\bibitem{Plank}
M. J. Plank and R. Law, Spatial point processes and moment dynamics in the life sciences: a parsimonious derivation and some extensions, \textit{Bull. Math. Biol.} \textbf{77} (2015) 586--613. 

\bibitem{Binny}
R. N. Binny, M. J. Plank and A. James, Spatial moment dynamics for collective cell movement incorporating a neighbour-dependent directional bias, \textit{J. Royal Soc. Interface} \textbf{12} (2015) 20150228. 

\bibitem{Binnya}
R. N. Binny, P. Haridas, A. James, R. Law, M. J. Simpson and M. J. Plank, Spatial structure arising from neighbour-dependent bias in collective cell movement, \textit{PeerJ} \textbf{4} (2016) e1689. 


\bibitem{FinKonKoz}
D. Finkelshtein, Y. Kondratiev and Y. Kozitsky, The statistical dynamics of a spatial logistic model and the related kinetic equation, \textit{Math. Models Meth. Appl. Sci.} \textbf{25} (2015) 343--370. 


\bibitem{Ruelle}
D. Ruelle, Superstable interactions in classical statistical mechanics, \textit{Commun. Math. Phys.} \textbf{18} (1970) 127--159. 


\bibitem{Young}
W. R. Young, A. J. Roberts and G. Stuhne, Reproductive pair correlations and the clustering of organisms, \textit{Nature} \textbf{412} (2001) 328--331. 


\bibitem{LawIll}
R. Law, J. Illian, D. F. R. P. Burslem, G. Gratzer, C. V. S. Gunatilleke and I. A. U. N. Gunatilleke, Ecological information from spatial patterns of plants: insights from point process theory, \textit{J. Ecol.} \textbf{97} (2009) 616--628. 

\bibitem{Agnew}
D. J. G. Agnew, J. E. F. Green, T. M. Brown, M. J. Simpson, and B. J. Binder, Distinguishing between mechanisms of cell aggregation using pair-correlation functions, \textit{J. Theor. Biol.} \textbf{352} (2014) 16--23. 


\bibitem{Fuentes}
M. A. Fuentes, M. N. Kuperman and V. M. Kenkre, Nonlocal interaction effects on pattern formation in population dynamics, \textit{Phys. Rev. Lett.} \textbf{91} (2003) 158104. 


\bibitem{Fisher}
R. A. Fisher, The wave of advance of advantageous genes, \textit{Ann. Eugen.} \textbf{7} (1937) 355--369. 

\bibitem{Kolomogor}
A. N. Kolmogoroff, I.G. Petrovsky and N. S. Piscounoff, \'Etude de l\'equation de la diffusion avec croissance de la quantit\'e de mati\'ere et son application \'a un probl\'eme biologique, \textit{Moscow Univ. Bull. Math.} \textbf{1} (1937) 1--25.

\bibitem{Maruvka}
Y. E. Maruvka and N. M. Shnerb, Nonlocal competition and logistic growth: Patterns, defects, and fronts, \textit{Phys. Rev. E} \textbf{73} (2006) 011903. 


\bibitem{Baker}
R. E. Baker and M. J. Simpson, Correcting mean-field approximations for birth-death-movement processes, \textit{Phys. Rev. E} \textbf{82} (2010) 041905. 

\bibitem{Markham}
D. C. Markham, M. J. Simpson and R. E. Baker, Simplified method for including spatial correlations in mean-field approximations, \textit{Phys. Rev. E} \textbf{87} (2013) 062702. 

\bibitem{Markhama}
D. C. Markham, M. J. Simpson, P. K. Maini, E. A. Gaffney and R. E. Baker, Incorporating spatial correlations into multispecies mean-field models, \textit{Phys. Rev. E} \textbf{88} (2013) 052713. 

\bibitem{Simpsona}
M. J. Simpson and R. E. Baker, Corrected mean-field models for spatially dependent advection-diffusion-reaction phenomena, \textit{Phys. Rev. E} \textbf{83} (2011) 051922. 


\bibitem{Kirkwood}
J. G. Kirkwood, Statistical mechanics of fluid mixtures, \textit{J. Chem. Phys.} \textbf{3} (1935) 300--313. 

\bibitem{Hansen}
J.-P. Hansen and I. McDonald, \textit{Theory of Simple Liquids} (Elsevier Academic Press, London, 2006), 3rd edition. 

\bibitem{BenNaim}
A. Ben-Naim, The Kirkwood superposition approximation, revisited and reexamined, \textit{Journal of Advances in Chemistry} \textbf{1} (2013) 27--35. 


\bibitem{Omelyans}
I. P. Omelyan, I. M. Mryglod and R. Folk, Algorithm for molecular dynamics simulations of spin liquids, \textit{Phys. Rev. Lett.} \textbf{86} (2001) 898--901. 

\bibitem{Omelyana}
I. P. Omelyan, I. M. Mryglod and R. Folk, Symplectic analytically integrable decomposition algorithms: classification, derivation, and application to molecular dynamics, quantum and celestial mechanics simulations, \textit{Comput. Phys. Commun.} \textbf{151} (2003) 272--314. 


\bibitem{Smaldino}
P. E. Smaldino, J. C. Schank and R. McElreath, Increased costs of cooperation help cooperators in the long run, \textit{Am. Nat.} \textbf{181} (2013) 451--463. 

\bibitem{Demirel}
G. Demirel, F. Vazqueza, G. A. B\"{o}hmea and T. Gross, Moment-closure approximations for discrete adaptive networks, \textit{Physica D} \textbf{267} (2014) 68--80. 


\bibitem{Somani}
S. Somani, B. J. Killian and M. K. Gilson, Sampling conformations in high dimensions using low-dimensional distribution functions, \textit{J. Chem. Phys.} \textbf{130} (2009) 134102. 

\end{thebibliography}
\end{document}